\documentclass[aps,pra,reprint,aps,twocolumn,superscriptaddress,showpacs,showkeys]{revtex4-1}

\usepackage{amsmath,amsthm,amssymb,graphicx,xcolor,float}

\usepackage{esint}
\usepackage[unicode=true]{hyperref}
\hypersetup{
     colorlinks=true,       		
     linkcolor=blue,          	
     citecolor=red,            
     urlcolor=magenta,           	
 }

\usepackage{mathrsfs,mathtools,mathdots,dsfont}
\usepackage{listings}
\usepackage{subfigure}

\newtheorem*{theorem*}{Theorem}

\newtheorem*{corollary*}{Corollary}

\newtheorem*{lemma*}{Lemma}

\newtheorem*{proposition*}{Example*}

\newtheorem*{conjecture*}{Conjecture}
\theoremstyle{definition}

\newtheorem*{definition*}{Definition}
\theoremstyle{remark}
\newtheorem{remark}{Remark}
\newtheorem*{remark*}{Remark}

\newcommand{\e}{\mathrm{e}}
\renewcommand{\i}{\mathrm{i}}

\newcommand{\ket}[1]{\left|#1\right\rangle}
\newcommand{\bra}[1]{\left\langle#1\right|}
\newcommand{\Eq}[1]{{\rm Eq.}~\eqref{#1}}

\begin{document}

\title{Goos-H{\"a}nchen Shift for Relativistic Particles Based on Dirac's Equation}

\author{Jiang-Lin Zhou}
\affiliation{School of Physics, Nankai University, Tianjin 300071, People's Republic of China}

\author{Zhen-Xiao Zhang}
\affiliation{School of Physics, Nankai University, Tianjin 300071, People's Republic of China}

\author{Xing-Yan Fan}
\affiliation{Theoretical Physics Division, Chern Institute of Mathematics, Nankai University, Tianjin 300071, People's Republic of
	 	China}

\author{Jing-Ling Chen}
\email{chenjl@nankai.edu.cn}
\affiliation{Theoretical Physics Division, Chern Institute of Mathematics, Nankai University, Tianjin 300071, People's Republic of
	 	China}

\date{\today}

\begin{abstract}
    The Goos-H{\"a}nchen (GH) shift is a specifical optical phenomenon that describes a shift parallel to the reflected light inside the plane of incidence,  when a finite-width light undergoes total
    internal reflection at the interface of medium. Although the GH shift in optics has been widely observed experimentally, its generalization remains uncovered completely in relativistic quantum mechanics for the existence of Klein's paradox. Recently, Wang has solved Klein's paradox based on the different solutions adpoted for Dirac's equation with step potential in corresponding energy regions \href{https://dx.doi.org/10.1088/2399-6528/abd340}{[J. Phys. Commun. {\bf 4}, 125010 (2020)]}. In the light of Wang's method, we calculate the GH shift for Dirac fermions under relativistic conditions when they are incident obliquely on a three-dimensional infinite potential barrier. Furthermore, we find that the relativistic quantum GH shift can be negative, which is different from the non-relativistic case.
\end{abstract}

\maketitle

\section{Introduction}
Retrospect to Newton, who proposed that when a finite-sized beam of light totally reflects from a glass interface
into a vacuum, photons first enter the vacuum before being pulled back into the glass medium, resulting in a lateral shift \cite{newton1730or}. However, this phenomenon was not observed experimentally until 1947 \cite{1947Goos}, which was then named the Goos-H{\"a}nchen (GH) shift. Further they conducted systematic observations of the displacement phenomenon \cite{1949Goos}. Since then, research on the GH shift has continued to develop. Later Artmann analyzed
the phenomenon theoretically \cite{1948Artmann}, where he used the stationary phase method to consider that a finite-width beam of light could be regarded as a composition of many plane waves; based on the assumption that the reflected beam should be considered as a whole plane wave, so it is necessary to maintain phase consistency at each wavefront; this ultimately
leads to the expression of GH shift
\begin{figure}[t]
    \includegraphics[width=0.58\textwidth]{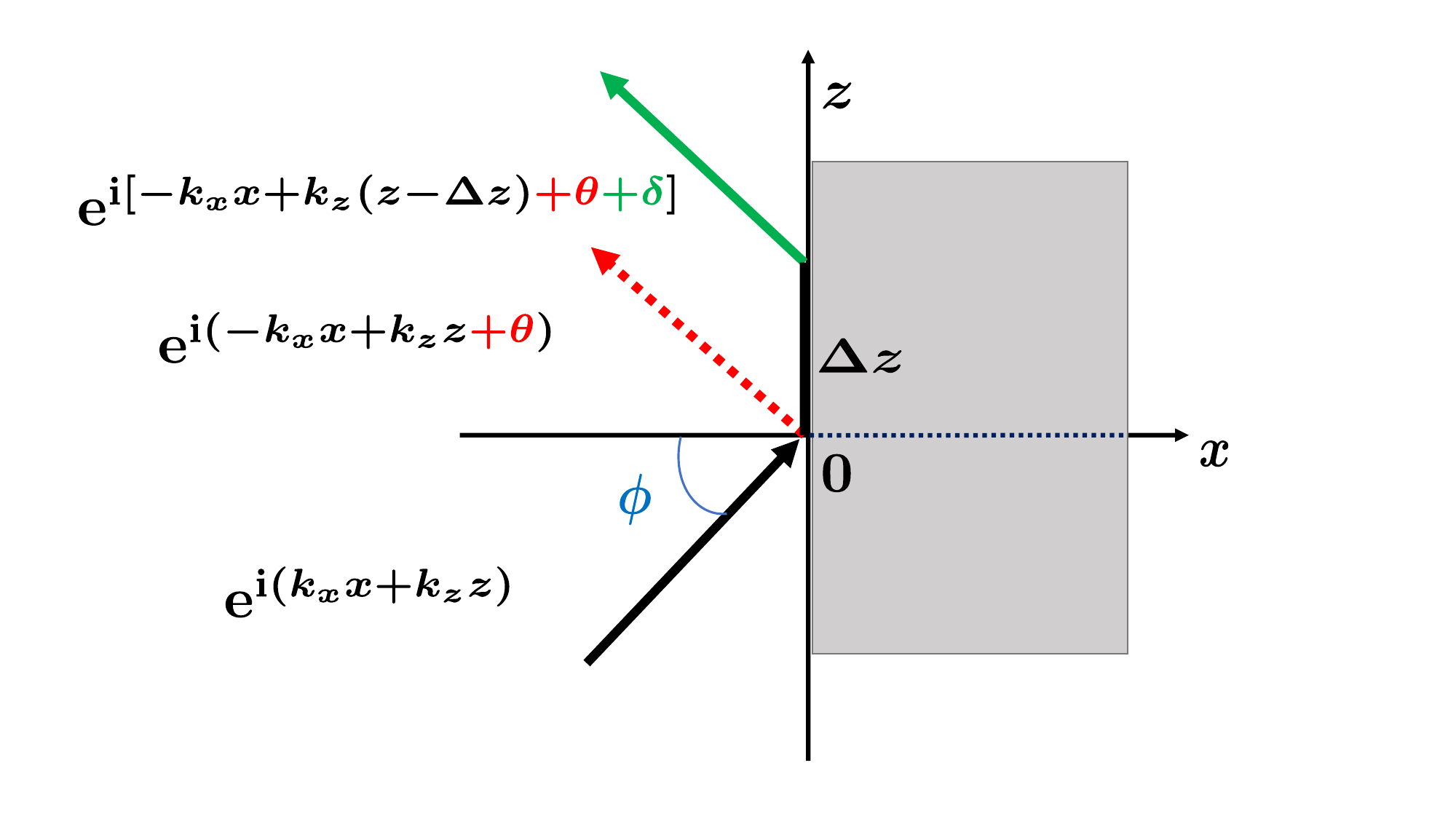}
    \caption{{\bf The diagram of optical GH shift}. Suppose the wave expression of incident beam (black solid) is $\psi(x,z)=\e^{\i (k_x x+k_z z)}$, with the incident angle $\phi$; where $k_x$, $k_z$ are the components of wave vector $\vec{k}$ along $x$-axis and $z$-axis respectively. The red dashed line is the reflected beam with the GH shift $\Delta z$ neglected, as well as the phase term $\theta$ obtained after reflection. If one considers the GH shift, the red dashed line changes into the green solid one, together with a new phase shift $\delta$.}\label{fig:OptGHShift}
\end{figure}
\begin{equation}
\Delta z=-\frac{\partial\, \theta}{\partial k_{z}},
\end{equation}
as shown in FIG. \ref{fig:OptGHShift}, where $k_z$ presents the $z$-component of wave vector $\vec{k}$, and $\theta$ expresses the phase obtained after reflection.

Remarkably, akin to the GH shift, there exists another effect dubbed Imbert-Fedorov shift \cite{fedorov1955k,imbert1972calculation} perpendicular to the plane of incidence, and all of these two effects have been unified using the approach of matrix optics \cite{li2007unified}, which is very similar to the method used to analyze the GH shift under quantum circumstances.

Contrary to Newton's initial assumption, the GH shift is a purely wave phenomenon \cite{widelytalk}, and thus should also be observable in quantum mechanics. In 1981, A. A. Seregin gave an analytical result about the GH shift of a neutron \cite{seregin1981surface}. Then, further analysis and many possible experimental plans were also proposed \cite{maaza1997possibility,ignatovich2004neutron}. However, since it is on the order of the wavelength of the incident wave \cite{zhuqibiao}, the GH shift of the particle wavefunction is almost impossible to be observed directly. The experimental breakthrough appeared in 2010, when V.-O. de Haan and the collaborators indirectly observed the GH shift during neutron scattering \cite{de2010observation}. They used the neutrons with spin-polarized current, reflected on the surface of a magnetized thin film. The existence of GH shift was determined by measuring the slight change in the direction of neutron polarization caused by the deviation of the GH shift of the upper and lower wavefunctions in the neutron spin representation, due to the corresponding energy differences in the potential barrier. Undoubtedly, their experiment provided a solid foundation for theoretical research on quantum GH shift.

In fact, due to the aforementioned observational difficulties, research on the GH shift of the particle wavefunction has long been mainly theoretical in distinct situations, such as massless Dirac fermions on the surface of graphene \cite{beenakker2009quantum}, large mass fermions (electrons) in bilayer graphene under ``step potential barrier'' conditions \cite{cheng2012goos}, and cold atoms with spin angular momentum as well as orbital angular momentum coupling \cite{zhou2015goos}. There are even more complicated cases, including the study of the phenomenon under the action of time-periodic potentials  \cite{huaman2019anomalous}, massless carriers in semi-Dirac materials encountering electric potential barriers during transport \cite{xiang2024total}, and the relationship between the GH shift occurring on the surface of Weyl semimetals with the properties thereof \cite{ye2019goos,dongre2022effects}.
What is worthy of a special mention is that X. Chen \emph{et al.} made a detailed discussion on various practical situations of electron's GH shift \cite{chen2013electronic}.

However, so far, all the research on the quantum GH shift has been conducted under the non-relativistic conditions. This is due to the existence of the famous Klein paradox \cite{klein1929reflexion}: the reflection coefficients may exceed one, when the Dirac fermions are incident on a one-dimensional infinite potential barrier under high-energy conditions \cite{yndurain2012relativistic}. To address the anomalous reflection coefficient, Dirac gave his equally famous explanation of the ``Dirac sea'' \cite{1928Dirac}, but this explanation is still defective, such as being unable to explain the existence of anti-bosons. Various explanations have been proposed for Klein's paradox, but none of them is able to reasonably explain the problem of reflection coefficients, thus making the calculation of the GH shift in higher dimensions meaningless \cite{1999Calogeracos}. It is worth mentioning that in 2020, Wang gave a relatively reasonable explanation for Klein's paradox by rationally choosing from positive and negative energy solutions under different circumstances of energy and potential barrier \cite{wang2020solving}.

The remainder of this paper will show
how we use Wang's treatment for Klein's paradox to deal with the situation of Dirac fermions obliquely incident on a three-dimensional infinite
potential barrier, obtaining reasonable reflection and transmission coefficients in Sec. \ref{sec:1DOblqIncid}. Relativistic quantum GH shift is calculated in Sec. \ref{sec:RQGHShift}. Discussion about the positive and negative GH shifts is given in Sec.\ref{sec:Dis}. Conclusion is made in 
Sec. \ref{sec:Conc}.

\section{Oblique Incidence of Dirac's Particle}\label{sec:1DOblqIncid}

In this section, we conider the oblique incidence of Dirac's particle. For simplicity, we consider the incident particle with momentum components only in the $x$ and $z$ directions, i.e., $\vec{p}=(p_x,0,p_z)$, as depicted in FIG. \ref{fig:OptGHShift}.

It is well known that Dirac's equation is a fundamental equation in the relativistic quantum mechanics. For a free Dirac's particle, its Hamiltonian is given by $H=(\vec{\alpha}\cdot\vec{p})c-\beta\,m\,c^2$ and its corresponding energy reads $E_\pm=\pm \sqrt{p^2 c^2 +m^2 c^4}$, where $m$ is the static mass, $c$ is the speed of light in vacuum, $
    \vec{\alpha}\equiv\sigma_x \otimes\vec{\sigma}$ and $
    \beta\equiv\sigma_z \otimes\openone$ are Dirac's matrices,  $
    \mathbb{I}\equiv\openone \otimes \openone$,  $\vec{\sigma}=(\sigma_x,\sigma_y,\sigma_z)$ depicts the vector of Pauli operators, and $\openone$ refers to the two-dimensional identity operator. Let $\ket{\psi}$ be the wavefunction of a free Dirac's particle, due to $H\ket{\psi}=E\ket{\psi}$, then we can express the stationary Dirac equation $\big[E\,\mathbb{I}-(\vec{\alpha}\cdot\vec{p})c-\beta\,m\,c^2\big]\ket{\psi}=0$ in a matrix form as
    \begin{equation}
    \begin{pmatrix}
        E - mc^2 & 0 & -p_z c & -p_x c \\
        0 & E - mc^2 & -p_x c & p_z c \\
        -p_z c & -p_x c & E + mc^2 & 0 \\
        -p_x c & p_z c & 0 & E + mc^2
    \end{pmatrix}\ket{\psi}= 0,
\end{equation}
from which we derive the general non-normalized solutions $\ket{\psi_\pm}$ with positive/negative energies as follows:
\begin{align}
\ket{\psi_+} =& f_1 \begin{pmatrix}
        E + mc^2 \\
        0 \\
        p_z c \\
        p_x c
    \end{pmatrix}+ f_2 \begin{pmatrix}
        0 \\
        E + mc^2 \\
        p_x c \\
        -p_z c
    \end{pmatrix}, \\
\ket{\psi_-} =& f_3 \begin{pmatrix}
        -E + mc^2 \\
        0 \\
        p_z c \\
        p_x c
    \end{pmatrix}+ f_4 \begin{pmatrix}
        0 \\
        -E + mc^2 \\
        p_x c \\
        -p_z c
    \end{pmatrix},
\end{align}
where $f_1$, $f_2$, $f_3$ and $f_4$ are four coefficients of superposition. Further if we ponder the temporal and spacial propagation term, a possible full form of wavefunction with positive energy shapes
\begin{align}
    \ket{\Psi_+} =& \left[\begin{pmatrix}
        E + mc^2 \\
        0 \\
        p_z c \\
        p_x c
    \end{pmatrix}+  \ell\begin{pmatrix}
        0 \\
        E + mc^2 \\
        p_x c \\
        -p_z c
    \end{pmatrix}\right]\e^{\frac{\i}{\hbar} (p_x x+p_z z-E\,t)} \notag \\
    =& \begin{pmatrix}
        E + mc^2 \\
        \ell(E + mc^2) \\
        (p_z +\ell\,p_x) c \\
        (p_x -\ell\,p_z)c
    \end{pmatrix}\e^{\frac{\i}{\hbar} (p_x x+p_z z-E\,t)},
\end{align}
which means $f_1=\e^{(\i/\hbar)(p_x x+p_z z-E\,t)}$, and $f_2/f_1=\ell$, with $\i^2 =-1$, $\hbar$ the reduced Planck constant, and $\ell$ an arbitrary constant. Similarly, one may have the wavefunction $\ket{\Psi_-(x, z, t)}$ with negative energy.

When this particle encounters an infinitely wide potential barrier with the potential $V(x)$, described by
\begin{equation}
    V(x) = \begin{cases}
        V, & x > 0, \\
        0, & x \leq 0,
    \end{cases}
\end{equation}
with $V>0$. Dirac's equation in the representation of position now becomes
\begin{equation}
\begin{cases}
    \bigg(\i \hbar\dfrac{\partial}{\partial t} -V
        +\i \hbar\,c\,\vec{\alpha}\cdot \vec{\nabla} - \beta\,m\,c^2
    \bigg)\ket{\Psi}=0, & x > 0, \\
    & \\
    \bigg(\i \hbar\dfrac{\partial}{\partial t}
        +\i \hbar\,c\,\vec{\alpha}\cdot \vec{\nabla} - \beta\,m\,c^2
    \bigg)\ket{\Psi}=0, & x \le 0,
\end{cases}
\end{equation}
with the operator $\vec{\nabla}\equiv(\partial/\partial x,\partial/\partial y,\partial/\partial z)$ and $\vec{p}=-{\rm i}\hbar \vec{\nabla}$. If one focuses on the scenario of \( V > E - \sqrt{m^2c^4 +p_z^2c^2} \), where quantum tunneling occurs, then one will encounter the so-called Klein's Paradox. In accordance with Wang's idea of solving Klein's Paradox \cite{wang2020solving}, we denote the wavefunction at the incident end as the positive energy solution, while as the negative energy solution at the transmission end. Thus, we write the wavefunction in the following way:
\onecolumngrid
\begin{align}
\ket{\Psi_1}=\ket{\Psi_{\rm in}}+\ket{\Psi_{\rm rf}}\equiv\begin{pmatrix}
        E + mc^2 \\
        \ell(E + mc^2) \\
        (p_z + \ell\,p_x)c \\
        (p_x - \ell\,p_z)c
    \end{pmatrix}\e^{\frac{\i}{\hbar} (p_x x + p_z z - Et)}
    + \left\{A \begin{pmatrix}
        E + mc^2 \\
        0 \\
        p_z c \\
        -p_x c
    \end{pmatrix}
    + B \begin{pmatrix}
        0 \\
        E + mc^2 \\
        -p_x c \\
        -p_z c
    \end{pmatrix}\right\}{\rm e}^{\frac{\i}{\hbar} (-p_x x + p_z z - Et)},
\end{align}
and
\begin{equation}
\ket{\Psi_2} = \left\{C \begin{pmatrix}
        V - E + mc^2 \\
        0 \\
        p_z c \\
        p_x' c
    \end{pmatrix}
    + D \begin{pmatrix}
        0 \\
        V - E + mc^2 \\
        p_x' c \\
        -p_z c
    \end{pmatrix}\right\} {\rm e}^{\frac{\i}{\hbar} (p_x' x + p_z z - Et)},
\end{equation}
\twocolumngrid
\noindent where $\ket{\Psi_1}$ and $\ket{\Psi_2}$ are the wavefunctions on the incident and transmitted half-plane, respectively; moreover $\ket{\Psi_1}$ is comprised of the incident wave $\ket{\Psi_{\rm in}}$ and the reflected wave $\ket{\Psi_{\rm rf}}$ with opposite $p_x$. The letters $\ell$, $A$, $B$, $C$, and $D$ are some coefficients. $\vec{p}=(p_x,0,p_z)$ and $\vec{p}{\, '}=(p'_x,0,p_z)$ are the momentum of the incident and transmitted particle, respectively. Notice that in the region of $x\leq 0$, the mass-energy relation reads
\begin{equation}\label{eq:E1}
    E^2 =p^2 c^2 +m^2 c^4 =(p_x^2 +p_z^2)c^2 +m^2 c^4,
\end{equation}
and in the region of $x>0$, the mass-energy relation becomes
\begin{equation}\label{eq:EV}
 (E-V)^2 =p'^2 c^2 +m^2 c^4 =(p_x'^2 +p_z^2)c^2 +m^2 c^4.
\end{equation}
 The $z$-components of all these three waves $\ket{\Psi_{\rm in}}$, $\ket{\Psi_{\rm rf}}$, and $\ket{\Psi_2}$ are identical for guaranteeing the continuity of phase on the surface of potential barrier. According to the continuity conditions at \( x = 0 \), we have
\begin{equation}\label{eq:ContinConds}
    \begin{cases}
    (E + mc^2)(1 + A) = C(V - E + mc^2), \\
    (E + mc^2)(\ell + B) = D(V - E + mc^2), \\
    p_z + \ell\,p_x + Ap_z- Bp_x= Cp_z + Dp_x', \\
    p_x - \ell\,p_z - Ap_x - Bp_z = Cp_x' - Dp_z.
    \end{cases}
\end{equation}
The coefficients $A$, $B$, $C$ and $D$ can be determined from Eq. (\ref{eq:ContinConds}) (see Appendix \ref{sec:DetalCoefs}).

According to the definition of reflection coefficient, one has the reflection coefficient as
\begin{equation}\label{eq:R1}
    R = \dfrac{|A|^2 + |B|^2}{1 + \ell^2},
\end{equation}
and the transmission coefficient as
\begin{equation}
    T = 1-R.
\end{equation}

By designating
\begin{equation}\label{eq:n}
    n = \frac{E + mc^2}{V - E + mc^2},
\end{equation}
for the region of  \( (E - V)^2 - p_z^2c^2 - m^2c^4 = p_x'^2 c^2 \ge 0 \), we have the reflection coefficient
\begin{align}\label{eq:RCoef}
R =& \dfrac{(p_x -n p_x')^2 +p_z^2 (1 - n)^2}{(p_x + n p_x')^2 +p_z^2 (1 - n)^2},
\end{align}
and the transmission coefficient as
\begin{align}\label{eq:TCoef}
T =& \dfrac{4\,n^2 p_x p_x' (V-E+m\,c^2)}{(E + mc^2)(np_x' + p_x)^2 +(n-1)^2 p_z^2}.
\end{align}
And for the region of  \( (E - V)^2 - p_z^2c^2 - m^2c^4 = p_x'^2 c^2 < 0 \), we have the total reflection coefficient and the zero transmission coefficient, i.e.,
\begin{equation}
            R = 1, \;\;\;  T = 0.
\end{equation}
The deduction on $R$ and $T$ are shown in Appendix \ref{sec:DetalCoefs}.

Through calculation, we find that \( R + T = 1 \) is valid for any value $V$ in the potential, thus successfully avoiding Klein's Paradox. When $(E - V)^2 - p_z^2c^2 - m^2c^4 = p_x'^2c^2 < 0 $, we find that \( p_x' \) becomes imaginary (i.e., $p_x'= {\rm i}\,q_x$), which results in the wavefunction at the transmission end resembling an evanescent wave.

\section{Relativistic Quantum GH Shift}\label{sec:RQGHShift}

As the cases in previous literatures, the GH shift occurs at the situation of total reflection. Thus, we turn our attention to the reflected wave within the potential range
\begin{equation}\label{eq:cond-1}
       (E - V)^2 - p_z^2c^2 - m^2c^4 < 0.
\end{equation}
Let is denote \( p_x' = \i\,q_x\), where
\begin{equation}
       q_x = \frac{1}{c}\sqrt{p_z^2 c^2 + m^2 c^4 - (V - E)^2}.
\end{equation}
 Subsequently, refer to the certain forms of $A$ and $B$ in Eqs. \eqref{eq:A} and \eqref{eq:B}, we may set
 \begin{equation}
    A=|A| {e}^{-\i\,\theta},B=|B| {e}^{-\i\,\theta}
\end{equation}
with $\theta$ the argument of the complex number
$[p_x^2 -n^2 q_x^2 + (1 - n)^2 p_z^2 ]+{\rm i}\,2n q_xp_x$, i.e.,
\begin{equation}
\tan \theta =\frac{2\,n\,p_x q_x} {p_x^2 -n^2 q_x^2 + (1 - n)^2 p_z^2}.
\end{equation}
Then the reflection wave function reads
\begin{align}
\ket{\Psi_{\rm r}} =\Bigg\{& |A|\begin{pmatrix}
        E + mc^2 \\
        0 \\
        p_z c \\
        -p_x c
    \end{pmatrix} \notag \\
    &+ |B| \begin{pmatrix}
        0 \\
        E + mc^2 \\
        -p_x c \\
        -p_z c
    \end{pmatrix}\Bigg\}\e^{
        -\i\,\theta+\frac{\i}{\hbar} (-p_x x + p_z z - Et)}.
\end{align}

We focus on the beam of infinite width before. Now, analogous to the GH Shift in optics, we hypothesize the existence of a GH shift for Dirac's particles. Considering a particle beam of finite width, similar to a light beam, we assume the beam consists of particles moving in different directions. Applying a solution akin to angular spectrum analysis in optics, we express the incident wavefunction as follows \cite{1948Artmann}
\begin{align}\label{incident wave fouriered}
& \ket{\psi(x, z)} \notag \\
=& \frac{1}{\sqrt{2 \pi}} \int_{-\infty}^{\infty}
    \ket{\psi_k(p_z)} \exp\bigg[\dfrac{\i}{\hbar} \big(p_z z + x \sqrt{p^2 - p_z^2}\big)\bigg] {\rm d} p_z,
\end{align}
where $\ket{\psi_k(p_z)}$ is the inverse Fourier transform of $
    \ket{\psi(x, z)}$. We define the average momentum of the beam as \( \vec{p}_0 = p_{x0} \,\vec{e}_x
    +p_{z0} \,\vec{e}_z \), and at \( x = 0 \),
\begin{equation}
\ket{\psi_k(p_z)}= \frac{1}{\hbar \sqrt{2 \pi}} \int_{-\infty}^{\infty}
    \ket{\psi(0, z)}\exp\left(-\frac{\i}{\hbar} p_z\,z\right){\rm d} z.
\end{equation}

Assume the probability of the particle appearing outside the beam width $2a$ is negligible.
Thus, we can express
\begin{equation}
    \ket{\psi(0, z)}= \begin{cases}
        \ket{\psi_0}\exp\left(\frac{\i}{\hbar} {p}_{z0}\,z\right), & |x| \leq a, \\
        0, & |x| > a,
    \end{cases}
\end{equation}
with
\begin{equation}\label{eq:Psi0}
    \ket{\psi_0}= \begin{pmatrix}
        E + mc^2 \\
        \ell(E + mc^2) \\
        (p_{z0} +\ell\,p_{x0})c \\
        (p_{x0} - \ell\,p_{z0})c
    \end{pmatrix}.
\end{equation}
It follows that
\begin{align}\label{eq:PsiKPz}
\ket{\psi_k(p_z)} =& \frac{1}{\hbar \sqrt{2 \pi}} \int_{-\infty}^{\infty}
    \ket{\psi(0, z)}\exp\left(-\frac{\i}{\hbar} p_z\,z
    \right){\rm d}z \notag \\
=& \frac{1}{\hbar \sqrt{2 \pi}} \ket{\psi_0}\int_{-a}^{a} \exp\left[
    \frac{\i}{\hbar} (p_{z0} -p_z)z\right]{\rm d}z \notag \\
=& \ket{\psi_0}\sqrt{\frac{2}{\pi}}\,\dfrac{\sin\left[\frac{(p_{z0} - p_z)a}{\hbar}\right]}{p_{z0} - p_z},
\end{align}
from which we conclude that the primary maximum occurs at \( p_z = p_{z0} \), and when \( a/\hbar \) is large, we only need to consider \( p_z \) near \( p_{z0} \).

\vspace{3mm}

\onecolumngrid
We introduce a reflection operator \( {\cal R} (p_z) \) to describe the behavior of the reflection wave
\begin{eqnarray}
\ket{\psi_{\rm r} (x, z)}=\frac{1}{\sqrt{2 \pi}}  \int_{-\infty}^{\infty} {\cal R} (p_z)\,
   \ket{\psi_k(p_z)}  \exp\left\{\dfrac{\i}{\hbar} \big[p_z z - x \sqrt{p^2 - p_z^2}
          - \hbar\,\theta(p_z)\big]\right\} {\rm d} p_z.
\end{eqnarray}
Given the minimal variation \( \delta p_z \), we focus solely on the wavefunction
\begin{equation}\label{reflection wave fouriered}
\ket{\psi_{\rm r} (x, z)}
= \dfrac{{\cal R}(p_{z0})}{\sqrt{2 \pi}} \int_{-\infty}^{\infty}
    \ket{\psi_k(p_z)}  \exp\left\{\dfrac{\i}{\hbar} \big[p_z z - x \sqrt{p^2 - p_z^2}
            - \hbar\,\theta(p_z)\big]\right\} {\rm d}p_z.
\end{equation}
Then we can solve the problem following Artmann's approach \cite{1948Artmann}, and we get the GH shift as
\begin{equation}\label{eq:DeltZ}
\Delta z=\frac{c\hbar  (E^2-m^2c^4-2EV+p^2 c^2 \cos2\phi) \,\tan\phi}{[-(E^2+m^2c^4)+p^2 c^2\cos2\phi] \sqrt{m^2c^4 - (E - V)^2 + p^2 c^2 \sin^2 \phi} },
\end{equation}
with $\phi\equiv\arctan(p_z /p_x)\in(0,\pi/2)$ indicating the incident angle. The deduction details are shown in Appendix \ref{sec:GHShf}.
 Note that $p_z=p \sin\phi$, thus the term inside the square root of \Eq{eq:DeltZ} is just $-[(E - V)^2 - p_z^2c^2 - m^2c^4]$, which is automatically positive.

\twocolumngrid

\begin{remark}
Analogous to the classical situation as shown in FIG. \ref{fig:OptGHShift}, there is a new phase shift $\delta$ attributed to the GH shift $\Delta z$, i.e.,
\begin{equation}
\delta=\frac{{p}_{z0}}{\hbar} \Delta z,
\end{equation}
whose deduction is given in Appendix \ref{sec:GHShf}.
\end{remark}


\section{Discussion: The Positive and Negative GH Shifts}\label{sec:Dis}

In this section we discuss the behavior of relativistic quantum GH shift $\Delta z$ in \Eq{eq:DeltZ} as a function of the incident angle $\phi$, energy $E$, and barrier height $V$, respectively. Bear in mind that the shift $\Delta z$ is calculated under the constraint of Eq. (\ref{eq:cond-1}), which is
\begin{equation}
(E - V)^2 - m^2c^4 - p^2 c^2 \sin^2 \phi < 0,
\end{equation}
i.e.,
\begin{equation}
(E - V)^2 - \left[m^2c^4 + (E^2-m^2c^4) \sin^2 \phi\right] < 0.
\end{equation}
Based on which, we have
\begin{equation}
 E-\gamma < V< E + \gamma,
\end{equation}
with
\begin{eqnarray}
 \gamma&=&\sqrt{m^2c^4 + (E^2-m^2c^4) \sin^2 \phi}\nonumber\\
 &=&\sqrt{ E^2 \sin^2 \phi + m^2c^4 \cos^2 \phi},
\end{eqnarray}
and $0\leq \gamma\leq E$.

\begin{figure}[t]
\centering
    \includegraphics[width=0.48\textwidth]{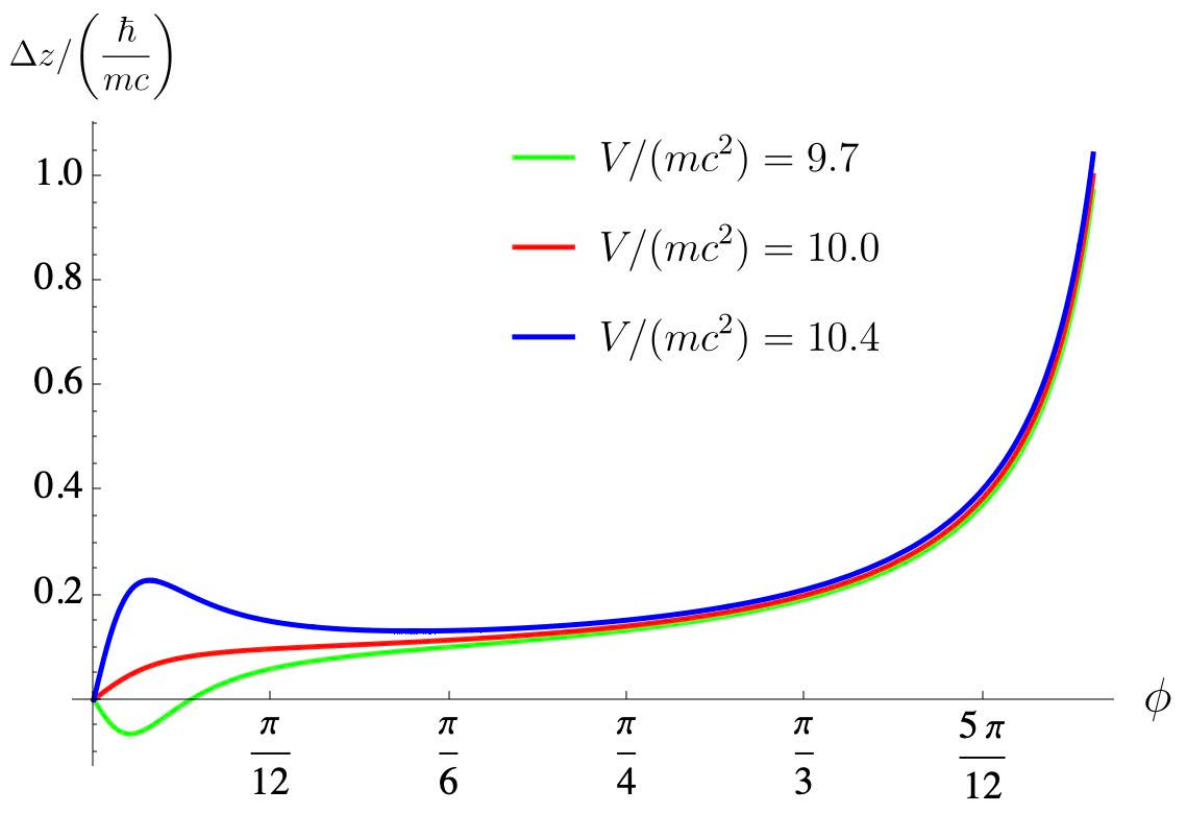}
        \caption{The behaviors of $\Delta z$ based on (\ref{eq:DeltZ-1a}) . (a). Fix $E=10\,m\,c^2$, the curve shows the change rule of $\Delta z$ with respect to the incident angle $\phi$ under three distinct potentials $V=9.7 mc^2$, $10.0 mc^2$, and $10.4 mc^2$.}
\label{fig2}
\end{figure}

With the help of
\begin{equation}
p^2 c^2=E^2-m^2c^4,
\end{equation}
Eq. (\ref{eq:cond-1}) can be recast to
\begin{eqnarray}\label{eq:DeltZ-1a}
\Delta z&=&\frac{c\hbar [(-E^2+m^2c^4)\cos^2 \phi+EV]\tan\phi }{\gamma^2 \sqrt{\gamma^2 -(E - V)^2 }},\
\end{eqnarray}
or
\begin{eqnarray}\label{eq:DeltZ-1b}
\Delta z &=&\frac{c\hbar [\gamma^2-(E-V)^2+(V-E)V]\tan\phi }{\gamma^2 \sqrt{\gamma^2 -(E - V)^2 }}.
\end{eqnarray}
The $\Delta z$-$\phi$ relation is shown in FIG. \ref{fig2}. We can have the results:

(i) From Eq. (\ref{eq:DeltZ-1a}) and  FIG. \ref{fig2} we find that there is no GH shift for $\phi=0$. Besides, greater the incident angle $\phi$ is, more similar the results with different potential are. Especially $\lim_{\phi\rightarrow\pi/2} \Delta z=\infty$, in view of the existence of the expression $\tan\phi$ in Eq. (\ref{eq:DeltZ-1a}).

(ii) Since $\gamma^2 -(E - V)^2\ge 0$, from Eq. (\ref{eq:DeltZ-1b}) we find that $\Delta z$ is always positive if $V\ge E$, which coincides with  FIG. \ref{fig2}.

(iii) $\Delta z$ can be negative, which means one has the negative GH shift. In fact, Beenakker \textit{et al.} have also discussed the negative GH shift in the situation of the massless Dirac fermions incident into graphene \cite{beenakker2009quantum}, and analogous results appeared in \cite{huaman2019anomalous,xiang2024total}.

Here we would like to discuss the occurrence of negative GH shift with the following two perspectives under the condition of $ V < E$.

(i). When $E$ and $\phi$ are fixed, from $\Delta z<0$, from Eq. (\ref{eq:DeltZ-1a}) we
have
    \begin{equation}\label{eq:evphi}
   (-E^2+m^2c^4)\cos^2 \phi+EV<0,
\end{equation}
which leads to
    \begin{equation}
    V<\left(E-\frac{m^2 c^4}{E}\right){\cos^2}\phi.
\end{equation}

(ii). When $E$ and $V$ are fixed, based on Eq. (\ref{eq:evphi}) the negative shift appears when
    \begin{equation}
  \cos^2 \phi< \frac{E^2-m^2c^4}{EV}.
\end{equation}

(iii) When $V$ and $\phi$ are fixed, based on Eq. (\ref{eq:evphi}) we have

    \begin{equation}
   E^2-\frac{V}{\cos^2 \phi}E -m^2c^4>0.
\end{equation}
By considering the incidence energy $E>0$, we obtain
    \begin{equation}
   E>\frac{V}{2 \cos^2 \phi}+\sqrt{\left(\frac{V}{2 \cos^2 \phi}\right)^2+m^2c^4}.
\end{equation}

\section{Conclusions}\label{sec:Conc}

In conclusion, we explore the oblique incidence of relativistic particles on an infinite potential barrier. To resolve the potential Klein paradox, we follow \cite{wang2020solving}, designating the wavefunction at the incident (transmitted) end as the positive (negative) energy solution.
Our calculations reveal that different potentials yield distinct outcomes. When \( (E - V)^2 - p_z^2 c^2 - m^2 c^4 <0 \), the $x$-component momentum of the transmittance particle \( p_x' \) becomes imaginary, akin to an evanescent wave. We then extend our discussion to the reflection wave, drawing parallels with the evanescent wave to postulate the existence of the GH shift. By analyzing a particle beam of finite width as well as conducting Fourier transformations, we derive the GH shift, and subsequently find the phenomenon of negative GH shift under specific configuration of $E$, $V$ and $\phi$.


Our theory is different with the non-relativistic result whose shift is always positive, and may inspire us
to explore the hide properties when the particle interacts with the partial well. Our theory also holds practical significance for realizing such as the design of neutron waveguides. However, due to the complexity of Dirac equation in the presence of an electromagnetic field, it needs more works to investigate the GH shift under that circumstance. Because the GH shift $\Delta z$ is so small that hard to be observed, the change of phase $\delta$ may be another choice to validate our theory.
\begin{acknowledgements}
J.L.C. is supported by the National Natural Science Foundation of China (Grants No. 12275136 and 12075001) and the 111 Project of B23045.

J.L.Z., Z.X.Z., and X.Y.F. contributed equally to this work.
\end{acknowledgements}

\onecolumngrid

\appendix

\section{Deduction of the Reflection Coefficient \eqref{eq:RCoef} and the Transmission Coefficient \eqref{eq:TCoef}}\label{sec:DetalCoefs}

Note after citing $n$ in \Eq{eq:n}, the equation set \eqref{eq:ContinConds} can be written in the following matrix form
\begin{equation}\label{eq:MCEqD}
M\,\vec{c}=\vec{d},
\end{equation}
where
\begin{equation}
M=\begin{pmatrix}
    n & 0 & -1 & 0\\
    0 & n & 0 & -1\\
    p_z & -p_x & -p_z & -p_x' \\
    p_x & p_z & p_x' & -p_z
\end{pmatrix},\qquad
\vec{c}=\begin{pmatrix}
    A\\
    B\\
    C\\
    D\\
\end{pmatrix},\qquad
\vec{d}=\begin{pmatrix}
    -n\\
    -\ell\,n\\
    -(p_z+\ell\,p_x) \\
    (p_x-\ell\,p_z)
\end{pmatrix}.
\end{equation}
Then we get
\begin{align}
\vec{c}=M^{-1} \vec{d}=\begin{pmatrix}
    \dfrac{p_x^2-n^2{p_x'}^2 - p_z^2(1 - n)^2 - 2\ell p_x p_z(1 - n)}{(np_x' + p_x)^2 + (1 - n)^2p_z^2}\\
    \dfrac{2(1 - n)p_xp_z + \ell \big[p_x^2 - n^2{p_x'}^2 - p_z^2(1 - n)^2\big]}{(np_x' + p_x)^2 + (1 - n)^2p_z^2}\\
    \dfrac{2np_x(p_x + np_x') - 2\ell np_xp_z(1 - n)}{(np_x' + p_x)^2 + (1 - n)^2p_z^2}\\
    \dfrac{2n(1 - n)p_xp_z + 2n\ell p_x(p_x + np_x')}{(np_x' + p_x)^2 + (1 - n)^2p_z^2}\\
\end{pmatrix}.
\end{align}
So we have
\begin{subequations}
\begin{align}
A =& \dfrac{p_x^2-n^2{p_x'}^2 - p_z^2(1 - n)^2 - 2\ell p_x p_z(1 - n)}{(np_x' + p_x)^2 + (1 - n)^2p_z^2}, \label{eq:A} \\
B =& \dfrac{2(1 - n)p_xp_z + \ell \big[p_x^2 - n^2{p_x'}^2 - p_z^2(1 - n)^2\big]}{(np_x' + p_x)^2 + (1 - n)^2p_z^2}, \label{eq:B} \\
C =& \dfrac{2np_x(p_x + np_x') - 2\ell np_xp_z(1 - n)}{(np_x' + p_x)^2 + (1 - n)^2p_z^2},
    \\
D =& \dfrac{2n(1 - n)p_xp_z + 2n\ell p_x(p_x + np_x')}{(np_x' + p_x)^2 + (1 - n)^2p_z^2}.
\end{align}
\end{subequations}

According to the $x$-component of the particle current density
\begin{equation}
j_x = c\bra{\Psi}\alpha_x \ket{\Psi},
\end{equation}
we arrive at
\begin{eqnarray}
j_{1x} &=&c\bra{\Psi_1}\alpha_x \ket{\Psi_1} =\left[2c(E + mc^2)p_x(1 + l^2)\right] -\left[2c(E + mc^2)p_x(|A|^2 + |B|^2)\right]\nonumber\\
&=& {\rm Term}_1-{\rm Term}_2,
\end{eqnarray}
and
\begin{equation}
j_{2x} =c\bra{\Psi_2}\alpha_x \ket{\Psi_2}=\begin{cases}
    2c(V - E + mc^2)p_x'(|C|^2 + |D|^2), & (E - V)^2 - p_z^2c^2 - m^2c^4 \geq  0, \\
    0, & (E - V)^2 - p_z^2c^2 - m^2c^4 <0.
\end{cases}
\end{equation}
From the definitions of reflection coefficient (i.e., the ratio ``${\rm Term}_2/{\rm Term}_1$'') and transmission coefficient (i.e., 
the ratio ``$j_{2x}/{\rm Term}_1$''), we have
\begin{equation}
R=\begin{cases}
    \dfrac{|A|^2 + |B|^2}{1 + l^2}, & (E - V)^2 - p_z^2c^2 - m^2c^4 \geq  0, \\
    1, & (E - V)^2 - p_z^2c^2 - m^2c^4 <0,
\end{cases}
\end{equation}
and
\begin{equation}
    T = \begin{cases}
        \dfrac{(V - E + mc^2)p_x'(|C|^2 + |D|^2)}{(E + mc^2)p_x(1 + l^2 )}, & (E - V)^2 - p_z^2c^2 - m^2c^4 \geq  0, \\
        0, & (E - V)^2 - p_z^2c^2 - m^2c^4 <0.
    \end{cases}
\end{equation}

When \( (E - V)^2 - p_z^2c^2 - m^2c^4 \ge0 \), the reflection and transmission coefficients are of the following forms:
\begin{align}
R =& \dfrac{|A|^2 + |B|^2}{1 + \ell^2}
=\dfrac{\Big[\frac{p_x^2-n^2{p_x'}^2 - p_z^2(1 - n)^2 - 2\ell p_x p_z(1 - n)}{
        (np_x' + p_x)^2 + (1 - n)^2p_z^2}\Big]^2
    +\Big\{\frac{2(1 - n)p_xp_z + \ell [p_x^2 - n^2{p_x'}^2 - p_z^2(1 - n)^2]}{
        (np_x' + p_x)^2 + (1 - n)^2p_z^2}\Big\}^2}{1 + \ell^2} \notag \\
=& \dfrac{(p_x -n p_x')^2 +p_z^2 (1 - n)^2}{(p_x + n p_x')^2 +p_z^2 (1 - n)^2}.
\end{align}
and
\begin{align}
T =& \dfrac{(V - E + mc^2)p_x'(|C|^2 + |D|^2)}{(E + mc^2)p_x(1 + \ell^2 )} \notag \\
=& \dfrac{(V - E + mc^2)p_x'\biggl\{
    \Big[\frac{2np_x(p_x + np_x') - 2\ell np_xp_z(1 - n)}{(np_x' + p_x)^2 + (1 - n)^2p_z^2}
    \Big]^2 +\Big[\frac{2n(1 - n)p_xp_z + 2\ell np_x(p_x + np_x')}{(np_x' + p_x)^2
        + (1 - n)^2p_z^2}\Big]^2\biggr\}}{(E + mc^2)p_x(1 + \ell^2 )} \\
=& \dfrac{4\,n^2 p_x p_x' (V-E+m\,c^2)}{(E + mc^2)(np_x' + p_x)^2 +(n-1)^2 p_z^2},
\end{align}
which satisfy $R+T=1$.

\section{Deduction of the GH Shift \eqref{eq:DeltZ}}\label{sec:GHShf}
From \Eq{eq:PsiKPz}, we let
\begin{align}
\ket{\psi_k (p_z)} =& \ket{\psi_0}\sqrt{\frac{2}{\pi}}\,
    \dfrac{\sin\left[\frac{(p_{z0} - p_z)a}{\hbar}\right]}{p_{z0} - p_z}
=\ket{\psi_0}\,F(p_z),
\end{align}
with
\begin{equation}
     F(p_z)=\sqrt{\frac{2}{\pi}} \frac{\sin\left[\frac{(p_{z0} - p_z)a}{\hbar}\right]}{p_{z0} - p_z},
\end{equation}
and $\ket{\psi_0}$ defined in \Eq{eq:Psi0}; then one arrives at
\begin{align}\label{eq:RWav}
\ket{\psi_{\rm r} (x, z)} =& \dfrac{{\cal R}(p_{z0})}{\sqrt{2 \pi}}
    \int_{-\infty}^{\infty} \ket{\psi_k(p_z)}\exp\left\{\dfrac{\i}{\hbar} \big[
        p_z z - x \sqrt{p^2 - p_z^2} - \hbar\,\theta(p_z)\big]\right\}{\rm d}p_z
    \notag \\
=& \dfrac{{\cal R}(p_{z0})}{\sqrt{2 \pi}}
    \int_{-\infty}^{\infty} \ket{\psi_0}\,F(p_z)\,\exp\Bigg(\dfrac{\i}{\hbar}
        \bigg\{p_z z - x \sqrt{p^2 - p_z^2} - \hbar\Big[\theta_0
            +\frac{\partial\,\theta}{\partial p_z} \Big|_{p_z = p_{z0}} (p_z - p_{z0})
                \Big]\bigg\}\Bigg){\rm d}p_z \notag \\
=& \dfrac{1}{\sqrt{2 \pi}} {\cal R}(p_{z0})\,\ket{\psi_0}\int_{-\infty}^{\infty} F(p_z)
    \,\exp\bigg(\dfrac{\i}{\hbar} \Big\{(\bar{p_z}+p_{z0})z - x \sqrt{p^2 - p_z^2} -\hbar\big[
        \theta_0 +\theta'_0 \bar{p_z}\big]\Big\}\bigg){\rm d}p_z \notag \\
=& \frac{1}{\sqrt{2 \pi}} \e^{\i(-\theta_0 +\frac{{p}_{z0} z}{\hbar})}
    {\cal R}(p_{z0}) \ket{\psi_0}\int_{-\infty}^{\infty} F(p_z)\,\exp\left\{
        \dfrac{\i}{\hbar} \big[\bar{p_z}(z -\hbar\,\theta'_0) - x\sqrt{p^2 - p_z^2}
            \big]\right\}{\rm d}p_z,
\end{align}
where
\begin{equation}
\theta(p_z) = \theta_0 + \theta'_0 \bar{p_z},\qquad
\theta'_0 \equiv\left.\dfrac{\partial\,\theta}{\partial p_z}\right|_{p_z = p_{z0}},
    \qquad
\bar{p_z}\equiv p_z - p_{z0}.
\end{equation}
\begin{remark}
Note the plane wave propagating along $(\pm)x$-axis has the following format
\begin{equation}
\ket{\psi(x, z)} = G\,\exp\left[\frac{\i}{\hbar} \big(\pm x \sqrt{p^2 - p_z^2}\big)
    \right]\int_{-\infty}^{\infty} F(p_z)\,\exp\left(\frac{\i}{\hbar} p_z\,z
        \right){\rm d}p_z,
\end{equation}
where $G$ is a coefficient, so the reflected wave \eqref{eq:RWav} behaves as a plane wave.
\end{remark}
Think about a wave parallel to the reflected wave with $\theta(p_z) = \theta_0$ (i.e., $\theta'_0 =0$) as follows:
\begin{equation}
\ket{\psi'(x, z)}\equiv\frac{1}{\sqrt{2 \pi}} \e^{
    \i(-\theta_0 +\frac{{p}_{z0} z}{\hbar})} {\cal R}(p_{z0}) \ket{\psi_0}
        \int_{-\infty}^{\infty} F(p_z)\,\exp\left[\dfrac{\i}{\hbar} \big(
            \bar{p_z}z- x\sqrt{p^2 - p_z^2}\big)\right]{\rm d}p_z,
\end{equation}
which coincides with the point of the incident wave at $x=0$ (no GH shift). Then we find
\begin{equation}
\e^{\i\,\delta}\ket{\psi'(x, z)}=\ket{\psi_{\rm r}(x, z +\Delta z)}.
\end{equation}
This shows that the reflected wave has a phase change $
    \delta\equiv(p_{z0} /\hbar)\Delta z$ and a shift $\Delta z$ along the $z$-axis with
\begin{align}
\Delta z &=\hbar\,\theta'_0
=\hbar\left.\dfrac{\partial\,\theta}{\partial p_z}\right|_{p_z = p_{z0}}
=\hbar\left.\dfrac{\partial}{\partial p_z} \Bigg(\arctan\biggl\{
    \dfrac{2\,n\,p_x q_x}{\big[p_x^2 -n^2 q_x^2 + (1 - n)^2 p_z^2\big]}\biggr\}
    \Bigg)\right|_{p_z = p_{z0}} \notag \\
=& \hbar\left.\dfrac{\partial}{\partial p_z} \Bigg(\arctan\biggl\{
    \dfrac{2\,n\,p_x \frac{\sqrt{p_z^2 c^2 + m^2 c^4 - (V - E)^2}}{c}}{
        p_x^2 -n^2 \frac{[p_z^2 c^2 + m^2 c^4 - (V - E)^2]}{c^2} + (1 - n)^2 p_z^2}
    \biggr\}\Bigg)\right|_{p_z = p_{z0}} \notag \\
=& \hbar\left.\dfrac{\partial}{\partial p_z} \Bigg(\arctan\biggl\{
    \dfrac{2\,n\,p_x c\sqrt{p_z^2 c^2 + m^2 c^4 - (V - E)^2}}{
        p_x^2 c^2 -n^2 [p_z^2 c^2 + m^2 c^4 - (V - E)^2]+(1 - n)^2 p_z^2 c^2}
    \biggr\}\Bigg)\right|_{p_z = p_{z0}} \notag \\
=& \hbar\left.\dfrac{\partial}{\partial p_z} \Bigg(\arctan\biggl\{
    \dfrac{2\,n\,p_x c\sqrt{p_z^2 c^2 + m^2 c^4 - (V - E)^2}}{
        p_x^2 c^2 -n^2 [m^2 c^4 - (V - E)^2]+(1 -2\,n)p_z^2 c^2}
    \biggr\}\Bigg)\right|_{p_z = p_{z0}} \notag \\
=& \hbar\left.\dfrac{\partial}{\partial p_z} \Bigg(\arctan\biggl\{
    \dfrac{2\,n\sqrt{E^2 -m^2 c^4 -c^2 p_z^2}\sqrt{m^2 c^4 - (V - E)^2 +c^2 p_z^2}}{
        E^2 -m^2 c^4 -n^2 [m^2 c^4 - (V - E)^2]-2\,n\,c^2 p_z^2}
    \biggr\}\Bigg)\right|_{p_z = p_{z0}} \notag \\
=& \dfrac{2\,c\,\hbar\,(E^2-m^2c^4-2EV+p^2 c^2 \cos2\phi)\tan\phi}{2[-(E^2+m^2c^4)+p^2 c^2\cos2\phi] \sqrt{m^2c^4 - V^2 + 2VE - E^2 + p^2 c^2 \sin^2 \phi}};
\end{align}
Here, \( p^2 c^2 = E^2 - m^2c^4 \) represents the momentum of the incident particle.






\twocolumngrid


\end{document}